\documentclass[a4paper]{article}
\usepackage[affil-it]{authblk}
\usepackage[dvipdfmx]{graphicx, color}
\usepackage{amssymb}
\usepackage{amsmath}
\usepackage{bm}
\usepackage{cases}
\usepackage{blkarray}
\usepackage{url}
\usepackage{color}
\usepackage{lineno}
\usepackage{cite}
\usepackage{setspace}

\begin{document}

\title{Comments on ``Y.-Q. Wang, et al., Commun. Nonlinear Sci. Numer. Simul. 103:105981 (2021)"}
\author[a]{Takahiro Ezaki\footnote{Correspondence: tkezaki@g.ecc.u-tokyo.ac.jp}}
\affil[a]{Research Center for Advanced Science and Technology, The University of Tokyo, 4-6-1 Komaba, Meguro-ku, Tokyo 153-8904, Japan}

\maketitle

\begin{abstract}
	With regard to the recently published article, ``Y.-Q. Wang, \textit{et al.}, Physical mechanism of equiprobable exclusion network with heterogeneous interactions in phase transitions: Analytical analyses of steady state evolving from initial state,  Commun. Nonlinear Sci. Numer. Simul. 103:105981 (2021)'', the following concerns are raised.
	First, the given steady-state solution of the system, its proof, and derivation of relevant physical quantities are incorrect.
	Second, the presented simulation results were not replicated and significantly deviated from their analytical predictions.
	Finally, the second part of the article is irrelevant in the context of equiprobable exclusion networks because the examined system in this part does not possess equiprobable properties.
\end{abstract}


(i) First, due to the fact that the detailed balance condition is not satisfied, Eqs. (1)--(4) in Ref. \cite{Wang2021-tl} are incorrect.
To prove the equations, the authors confirmed the detailed balance only for the transitions between subsystems in Eq. (5) in Ref. \cite{Wang2021-tl} but did not show the other types of transitions, for which reason the proof was incomplete.
In fact, the probability flows of the transitions of a particle between a subsystem (labeled by $j_1$) and the reservoir given the proposed solution Eqs. (1)--(4) do not cancel out in general as follows (see also Refs. \cite{Ezaki2012-bc,Ezaki2012-lg}):
\begin{eqnarray}
	&&
	P(\mathcal{C'})W(\mathcal{C'} \rightarrow \mathcal{C})
	- P(\mathcal{C})W(\mathcal{C} \rightarrow \mathcal{C'})
	\nonumber \\
	&&= \Xi ^{-1}\prod_{j=1, j\neq j_1}^{K} \left\{
	\left( \frac{w_{\rm a}}{w_j w_{\rm d}}\right)^{M_j}
	\left[
	\left(\frac{w_{\rm a}}{w_j w_{\rm d}}\right)^{M_{j_1}}  w_{\rm d}
	- \left(\frac{w_{\rm a}}{w_j w_{\rm d}}\right)^{M_{j_1} -1}  w_{\rm a}
	\right]
	\right\}\nonumber \\
	&& = \Xi ^{-1}\prod_{j=1, j\neq j_1}^{K} \left\{
	\left( \frac{w_{\rm a}}{w_j w_{\rm d}}\right)^{M_j}
	\left(\frac{w_{\rm a}}{w_j w_{\rm d}}\right)^{M_{j_1} -1}
	\left[
		\frac{w_{\rm a}}{w_j} -  w_{\rm a}
		\right]
	\right\} \neq 0,
\end{eqnarray}
where $M_j$ is the total number of particles in subsystem $j$, and $\mathcal{C}$ and $\mathcal{C'}$ are an arbitrary pair of particle configurations between which the system can transition by an attachment or detachment of a particle.
They cancel out if one sets $w_{\rm d}= w_j$ or $w_j = 1$ for all the subsystems. The former condition corresponds to the model in the previous study \cite{Ezaki2012-lg}, and the latter corresponds to a trivial case.
Also, one can confirm that the proof is incorrect in another way. If the detailed balance between any transitions between subsystems is satisfied, the solution does not depend on the topology of the network. This includes the case where no link is set in the system (namely, subsystems are separated from each other), given the ergodicity of the system. In such a case, the density of particles must depend only on the attachment and detachment rates (i.e., $w_{\rm a}$ and $w_{\rm b}$) while the equations presented in Ref. \cite{Wang2021-tl} explicitly depend on $w_j$.

Also, the use of Eqs. (6)--(10) to derive Eq. (11) in Ref. \cite{Wang2021-tl} is incorrect for this system.
The authors used the saddle point approximation, which is often used to evaluate the partition function when the number of particles in the system is conserved (as in Refs. \cite{Ezaki2011-pc, Ezaki2012-bc}).
Eq. (6) unnecessarily limits the number of particles to a single value.
However, in this case, because the number of particles is not conserved due to the presence of the reservoir, one can directly compute the partition function as in Ref. \cite{Ezaki2012-lg}.

\begin{figure*}[t]
	\centering\includegraphics[width=160mm]{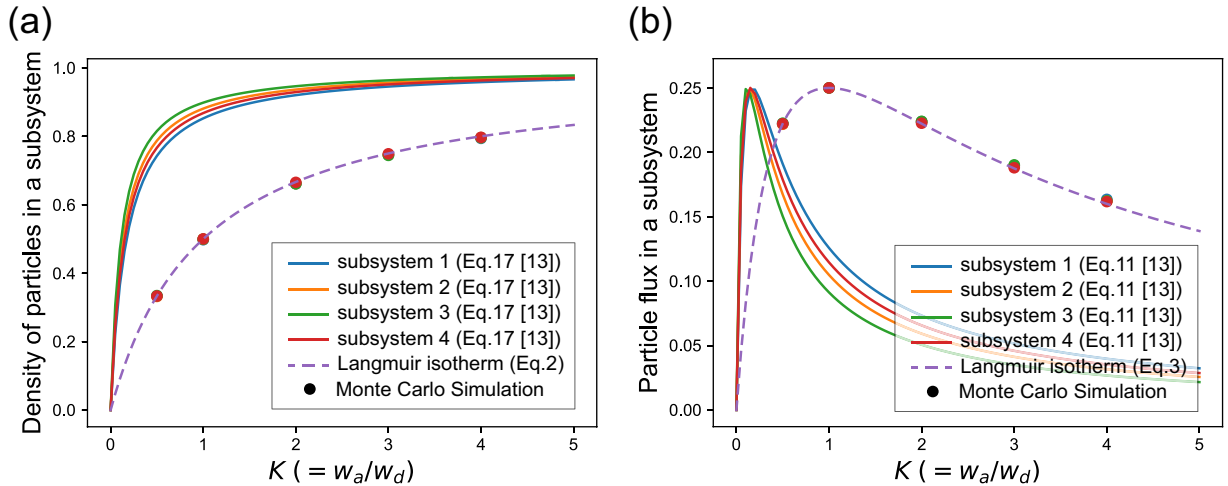}
	\caption{Comparisons between theory and simulation. (a) Density of particles in each subsystem. (b) Particle flux in each subsystem. The parameter values were the same as those used in Ref. \cite{Wang2021-tl}. As a network structure for the simulation (which information was not available in Ref. \cite{Wang2021-tl}), we randomly connected 50 pairs of sites bidirectionally. The solid lines represent the predictions by Wang \textit{et al.} \cite{Wang2021-tl}. The broken lines are the prediction by the Langmuir isotherm (Eqs. \eqref{eq:isotherm} and \eqref{eq:isotherm_J}).}
	\label{fig:comparison}
\end{figure*}

(ii) Wang \textit{et al.} \cite{Wang2021-tl} claimed that they verified the solutions by Monte Carlo simulations. However, the simulation results were not replicated for the density of particles (Fig. \ref{fig:comparison}(a); corresponding to Eq. (17) and Fig. 3 in Ref. \cite{Wang2021-tl}) nor the flux of particles (Fig. \ref{fig:comparison}(b); corresponding to Eq. (11) and Fig. 4 in Ref. \cite{Wang2021-tl}).
Instead, the results were approximately described by the Langmuir isotherm \cite{Ezaki2012-bc}:
\begin{equation}
	\rho_j = \frac{K}{1+K}, \: K=w_{\rm a}/w_{\rm b}, \label{eq:isotherm}
\end{equation},
\begin{equation}
	J_j = \frac{p_jK}{(1+K)^2} \label{eq:isotherm_J}.
\end{equation}
It should be also pointed out that the curve representing the theoretical prediction for subnetwork 4 and corresponding simulation results shown in Fig. 4 in Ref. \cite{Wang2021-tl} are incorrect. The flux of the ASEP (in continuous time) never exceeds 0.25, but it erroneously reached $\approx 0.33$ in the figure.
The same issue applies to Fig. 8 in Ref. \cite{Wang2021-tl}.
It is clear that the same conclusion is reached for Figs. 7--9 in Ref. \cite{Wang2021-tl}, which are the simple derivatives of $\rho_j$ and $J_j$.

(iii) Although the authors have investigated the dynamics of the ASEP with Langmuir kinetics in the latter half of the paper (Sec. 4 in Ref. \cite{Wang2021-tl}), because they assumed open boundary conditions in the model (Fig. 10 in the article) which breaks the equiprobale propertiy of the system, the results are completely irrelevant to the equiprobable network of the ASEPs, which was the main topic of the paper.

\end{document}